\documentclass[twocolumn,showpacs,aps,superscriptaddress]{revtex4-1}
\usepackage{amsmath}
\usepackage{graphicx}
\usepackage{dcolumn}
\usepackage{float}

\newcommand{\be}{\begin{equation}}
\newcommand{\ee}{\end{equation}}
\newcommand{\bea}{\begin{eqnarray}}
\newcommand{\eea}{\end{eqnarray}}
\newcommand{\bs}{\begin{split}}
\newcommand{\bse}{\begin{subequations}}
\newcommand{\ese}{\end{subequations}}

\begin{document}
\title{Investigations of the effect of nonmagnetic Ca substitution for magnetic Dy on spin-freezing in Dy$_2$Ti$_2$O$_7$}
\author{V. K. Anand}
\altaffiliation{vivekkranand@gmail.com}
\affiliation{\mbox{Helmholtz-Zentrum Berlin f\"{u}r Materialien und Energie, Hahn-Meitner Platz 1, D-14109 Berlin, Germany}}
\author{D. A. Tennant}
\altaffiliation{Present address: Neutron Sciences Directorate, Oak Ridge National Laboratory, Oak Ridge, TN 37831, U.S.A.}
\affiliation{\mbox{Helmholtz-Zentrum Berlin f\"{u}r Materialien und Energie, Hahn-Meitner Platz 1, D-14109 Berlin, Germany}}
\author{B. Lake}
\altaffiliation{bella.lake@helmholtz-berlin.de}
\affiliation{\mbox{Helmholtz-Zentrum Berlin f\"{u}r Materialien und Energie, Hahn-Meitner Platz 1, D-14109 Berlin, Germany}}
\date{\today}

\begin{abstract}
Physical properties of partially Ca substituted hole-doped Dy$_2$Ti$_2$O$_7$ have been investigated by ac magnetic susceptibility $\chi_{\rm ac}(T)$, dc magnetic susceptibility  $\chi(T)$, isothermal magnetization $M(H)$ and heat capacity $C_{\rm p}(T)$ measurements on Dy$_{1.8}$Ca$_{0.2}$Ti$_{2}$O$_{7}$. The spin-ice system Dy$_2$Ti$_2$O$_7$ exhibits a spin-glass type freezing behavior near 16~K\@. Our frequency dependent $\chi_{\rm ac}(T)$ data of Dy$_{1.8}$Ca$_{0.2}$Ti$_{2}$O$_{7}$ show that the spin-freezing behavior is significantly influenced by Ca substitution. The effect of partial nonmagnetic Ca$^{2+}$ substitution for magnetic Dy$^{3+}$ is similar to the previous study on nonmagnetic isovalent Y$^{3+}$ substituted Dy$_{2-x}$Y$_{x}$Ti$_{2}$O$_{7}$ (for low levels of dilution), however the suppression of spin-freezing behavior is substantially stronger for Ca than Y\@. The Cole-Cole plot analysis reveals semicircular character and a single relaxation mode in Dy$_{1.8}$Ca$_{0.2}$Ti$_{2}$O$_{7}$ as for Dy$_{2}$Ti$_{2}$O$_{7}$. No noticeable change in the insulating behavior of Dy$_2$Ti$_2$O$_7$ results from the holes produced by 10\% Ca$^{2+}$ substitution for Dy$^{3+}$ ions.
\end{abstract}

\pacs{75.50.Lk, 75.40.Gb, 75.40.Cx \\ Keywords: Spin freezing, Cole-Cole plot, Heat capacity, Magnetic susceptibility, Pyrochlore lattice}

\maketitle

\date{\today}

\section{\label{Intro} INTRODUCTION}

Rare earth pyrochlores $R_2B_2$O$_7$ ($R$ = rare earth and $B$ = transition metal) are known to present a wide range of interesting and unusual magnetic properties \cite{Gardner2010,Castelnovo2012,Harris1997,Ramirez1999,Siddharthan1999,Hertog2000,Bramwell2001,Castelnovo2008,Morris2009,Bramwell2009,Fennell2009}. The three-dimensional network of corner-sharing tetrahedra of $R$ ions in these cubic symmetry (space group $Fd\bar{3}m$) pyrochlores cause highly frustrated magnetism making them a model material among the geometrically frustrated spin systems. An interesting situation is encountered when the crystal electric field (CEF) forces the magnetic spin to point in a particular direction. The observation of spin-ice behavior in Dy$_2$Ti$_2$O$_7$ and Ho$_2$Ti$_2$O$_7$ is an exciting manifestation of such a CEF forced geometrically frustrated magnetism \cite{Harris1997,Ramirez1999,Bramwell2001} . In these spin-ice systems CEF anisotropy forces the magnetic spins at the corners of tetrahedra to align parallel to the tetrahedron axes (either towards the center or away from the center of tetrahedra). This causes an exotic magnetic state where even the ferromagnetic interaction between the Dy$^{3+}$ (Ho$^{3+}$) magnetic moments becomes frustrated  \cite{Harris1997}. The energy minimum state of magnetic exchange and dipolar interactions is achieved by two spins pointing inward and two spins outward on each tetrahedron, referred as the ``2-in \& 2-out" rule or ``ice rule" \cite{Harris1997}. The constraint condition on magnetic spins is analogous to the disordered protons in hexagonal water ice, accordingly the magnetic state of Dy$_2$Ti$_2$O$_7$ and Ho$_2$Ti$_2$O$_7$ is named as ``spin-ice". One of the major consequences of the ice rule is that a residual entropy remains at the lowest temperatures which is almost the same as the Pauling value for water ice  \cite{Ramirez1999}. Another striking consequence is the  realization of magnetic monopoles in spin-ice materials which interact through the magnetic Coulomb interaction \cite{Castelnovo2008}. The diffuse neutron scattering study on Dy$_2$Ti$_2$O$_7$ shows evidence of Dirac strings and magnetic monopoles \cite{Morris2009}. Polarized neutron scattering show evidence for a magnetic Coulomb phase in Ho$_2$Ti$_2$O$_7$ \cite{Fennell2009}. All these indicate a new state of magnetism in these materials which led to intense research activities on $R_2B_2$O$_7$ in recent years.

Apart from the spin-ice behavior below $T\sim 2$~K, the ac susceptibility measurements on Dy$_2$Ti$_2$O$_7$ also show a frequency-dependent anomaly near $T_f \approx 16$~K reflecting cooperative spin-freezing \cite{Snyder2001,Matsuhira2001}, however the dc susceptibility or heat capacity shows no corresponding anomaly. This spin-freezing feature near 16~K is not seen in spin-ice Ho$_2$Ti$_2$O$_7$ which distinguishes the spin-dynamics in Dy$_2$Ti$_2$O$_7$ and Ho$_2$Ti$_2$O$_7$. The spin-freezing behavior near $T_f$ in Dy$_2$Ti$_2$O$_7$ is distinct from the spin-freezing in spin-glasses. In a spin-glass system the application of a magnetic field is found to quench the spin-freezing. In contrast, in Dy$_2$Ti$_2$O$_7$ the application of magnetic field tends to enhance the spin-freezing \cite{Snyder2001}, thus indicating unusual dynamics of spin-freezing in Dy$_2$Ti$_2$O$_7$. We focus on this unusual glassiness in Dy$_2$Ti$_2$O$_7$ and study the effect on spin-freezing behavior near $T_f$ upon diluting the magnetic site and doping with holes, i.e. substituting nonmagnetic Ca$^{2+}$ for magnetic Dy$^{3+}$. The effect of dilution on spin-freezing behavior near 16~K has been previously investigated for nonmagnetic Y$^{3+}$ and Lu$^{3+}$ substitutions \cite{Snyder2002, Snyder2004}. The investigations on Dy$_{2-x}$Y$_{x}$Ti$_{2}$O$_{7}$ by Snyder {\it et al}. \cite{Snyder2002, Snyder2004} revealed that the 16~K freezing is initially suppressed for low dilution concentrations ($x\leq0.4$), however the freezing behavior was found to re-emerge for $x>0.4$, clearly reflecting the influence of local environment on the freezing process. The 16~K freezing in Dy$_2$Ti$_2$O$_7$ is considered to be an Orbach-like relaxation process which involves an excited CEF level for spin-lattice relaxations \cite{Orbach1961,Ehlers2003}. The spin dynamics near $T_f$ is controlled by transitions to/from the first excited doublet. Therefore the effect of dilution at Dy site seems to result from a change in CEF.

Here we report the physical properties of Dy$_{1.8}$Ca$_{0.2}$Ti$_{2}$O$_{7}$ based on ac magnetic susceptibility $\chi_{\rm ac}(T)$, dc magnetic susceptibility  $\chi(T)$, isothermal magnetization $M(H)$ and heat capacity $C_{\rm p}(T)$ measurements. The results are compared with those of Dy$_2$Ti$_2$O$_7$ and the effect of partial nonmagnetic Ca$^{2+}$ substitution for magnetic Dy$^{3+}$ on spin-freezing behavior near 16~K is addressed. The $\chi_{\rm ac}(T)$ data show the effect to be similar to that of nonmagnetic Y$^{3+}$ substitution, however the suppression of spin-freezing behavior is much stronger in the case of Ca substitution.

\section{\label{ExpDetails} EXPERIMENTAL DETAILS}
Polycrystalline samples of Dy$_{2-x}$Ca$_{x}$Ti$_{2}$O$_{7}$ ($x = 0$, 0.2) were synthesized by solid state reaction method. Dy$_2$Ti$_2$O$_7$ was synthesized using Dy$_2$O$_3$ (99.9\%, Sigma Aldrich) and TiO$_2$ (99.995\%, Alfa Aesar) in 1:2 molar ratio. The stoichiometric amounts were mixed and ground for half an hour using an agate mortar and pestle. The mixture was placed in a zirconium oxide crucible and fired thrice at 1350~$^{\circ}$C in air with an intermediate grinding every 50~h, after which the finely ground mixture was pressed into a pellet and fired at 1350~$^{\circ}$C for 50~h\@. For Dy$_{1.8}$Ca$_{0.2}$Ti$_{2}$O$_{7}$ synthesis Dy$_2$O$_3$, TiO$_2$ and CaCO$_3$ (99.95\%, Alfa Aesar) were taken in desired stoichiometry, mixed and ground thoroughly, and fired at 950~$^{\circ}$C for 50~h in air. The mixture was reground and fired at 1250~$^{\circ}$C for 50~h with two more subsequent intermediate grindings and firings at 1350~$^{\circ}$C for 50~h each and then the sample was pelletized and fired again at 1350~$^{\circ}$C for 50~h\@. A single phase sample of 10\% Ca substituted Dy$_{1.8}$Ca$_{0.2}$Ti$_{2}$O$_{7}$ could be obtained this way. However, similar synthesis did not yield single phase sample for 20\% or 30\% Ca substituted Dy$_{2-x}$Ca$_{x}$Ti$_{2}$O$_{7}$ or 10\% Na substituted Dy$_{1.8}$Na$_{0.2}$Ti$_{2}$O$_{7}$. A zirconium oxide crucible was used in each case.

Crystal structure and quality of sample were determined by powder x-ray diffraction (XRD).  Bulk properties were measured using facilities at Mag Lab, Helmholtz Zentrum Berlin. The dc magnetic measurements were carried out using a vibrating sample magnetometer superconducting quantum interference device (VSM-SQUID, Quantum Design) and the ac susceptibility was measured using the ACMS option of a physical properties measurement system (PPMS, Quantum Design). Heat capacity was measured by the adiabatic relaxation method using the heat capacity option of PPMS.

\section{\label{Results} Results and Discussion}

\begin{figure}
\begin{center}
\includegraphics[width=3.2in, keepaspectratio]{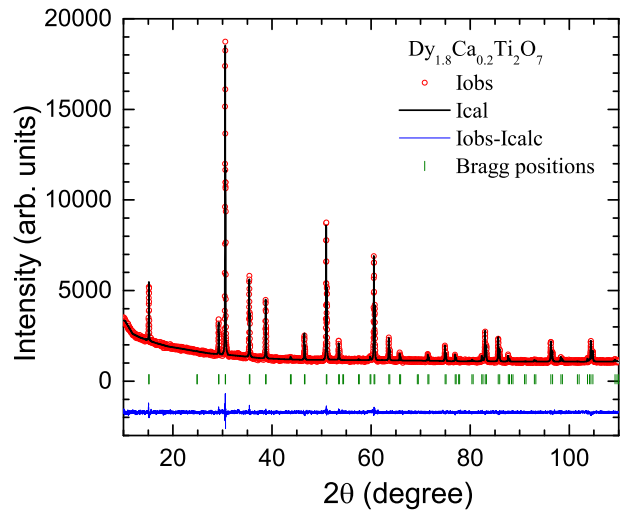}
\caption {(Color online) Powder x-ray diffraction pattern of Dy$_{1.8}$Ca$_{0.2}$Ti$_{2}$O$_{7}$ recorded at room temperature. The solid line through the experimental points is the Rietveld refinement profile calculated for the ${\rm Ca_2Nb_2O_7}$-type face centered cubic (space group $Fd\bar{3}m$) pyrochlore structure. The short vertical bars mark the Bragg peak positions. The lowermost curve represents the difference between the experimental and calculated intensities.}
\label{fig:XRD}
\end{center}
\end{figure}

\begin{table} 
\begin{center}
\caption{\label{tab:XRD} Crystallographic and refinement parameters obtained from the structural Rietveld refinement of powder XRD data of Dy$_2$Ti$_2$O$_7$ and Dy$_{1.8}$Ca$_{0.2}$Ti$_{2}$O$_{7}$. The atomic coordinates of Dy/Ca, Ti, O1 and O2 atoms in space group $Fd\bar{3}m$ are 16c (0,0,0), 16d (1/2,1/2,1/2), 48f ($x_{\rm O1}$,1/8,1/8) and 8a (1/8,1/8,1/8), respectively.}
\begin{ruledtabular}
\begin{tabular}{lcc}
 & Dy$_2$Ti$_2$O$_7$ & Dy$_{1.8}$Ca$_{0.2}$Ti$_{2}$O$_{7}$\\
 \hline
 \underline{Lattice parameters}\\
{\hspace{0.8cm} $a$ ({\AA})}            			&  10.1231(1) & 10.1311(1) \\	
{\hspace{0.8cm} $V_{\rm cell}$  ({\AA}$^{3}$)} 	&  1037.396(4)  & 1039.843(6) \\
\underline{Atomic coordinate}\\
\hspace{0.8cm} $x_{\rm O1}$ & 0.4243(5) & 0.4254(5)\\
\underline{Refinement quality} \\
\hspace{0.8cm} $\chi^2$   & 1.86 & 1.24\\	
\hspace{0.8cm} $R_{\rm p}$ (\%)  & 2.11 & 2.32 \\
\hspace{0.8cm} $R_{\rm wp}$ (\%) & 2.90 & 2.96 \\
\end{tabular}
\end{ruledtabular}
\end{center}
\end{table}

The powder XRD data were analysed by structural Rietveld refinement using the software {\tt FullProf} \cite{Rodriguez1993}. The room temperature XRD pattern and Rietveld refinement profile for Dy$_{1.8}$Ca$_{0.2}$Ti$_{2}$O$_{7}$ are shown in figure~\ref{fig:XRD}. The almost single phase nature of the sample is evident from the refinement of XRD data. The structural refinement shows that no stuctural instability is caused by 10\% Ca substitution for Dy, and like Dy$_2$Ti$_2$O$_7$, Dy$_{1.8}$Ca$_{0.2}$Ti$_{2}$O$_{7}$ also has ${\rm Ca_2Nb_2O_7}$-type face centered cubic (space group $Fd\bar{3}m$) pyrochlore structure. The crystallographic parameters of Dy$_{1.8}$Ca$_{0.2}$Ti$_{2}$O$_{7}$ are listed in Table~\ref{tab:XRD} and compared with those of Dy$_2$Ti$_2$O$_7$. The crystallographic parameters are in good agreement with the values reported for Dy$_2$Ti$_2$O$_7$ \cite{Fuentes2005}.

\begin{figure}
\begin{center}
\includegraphics[width=3in, keepaspectratio]{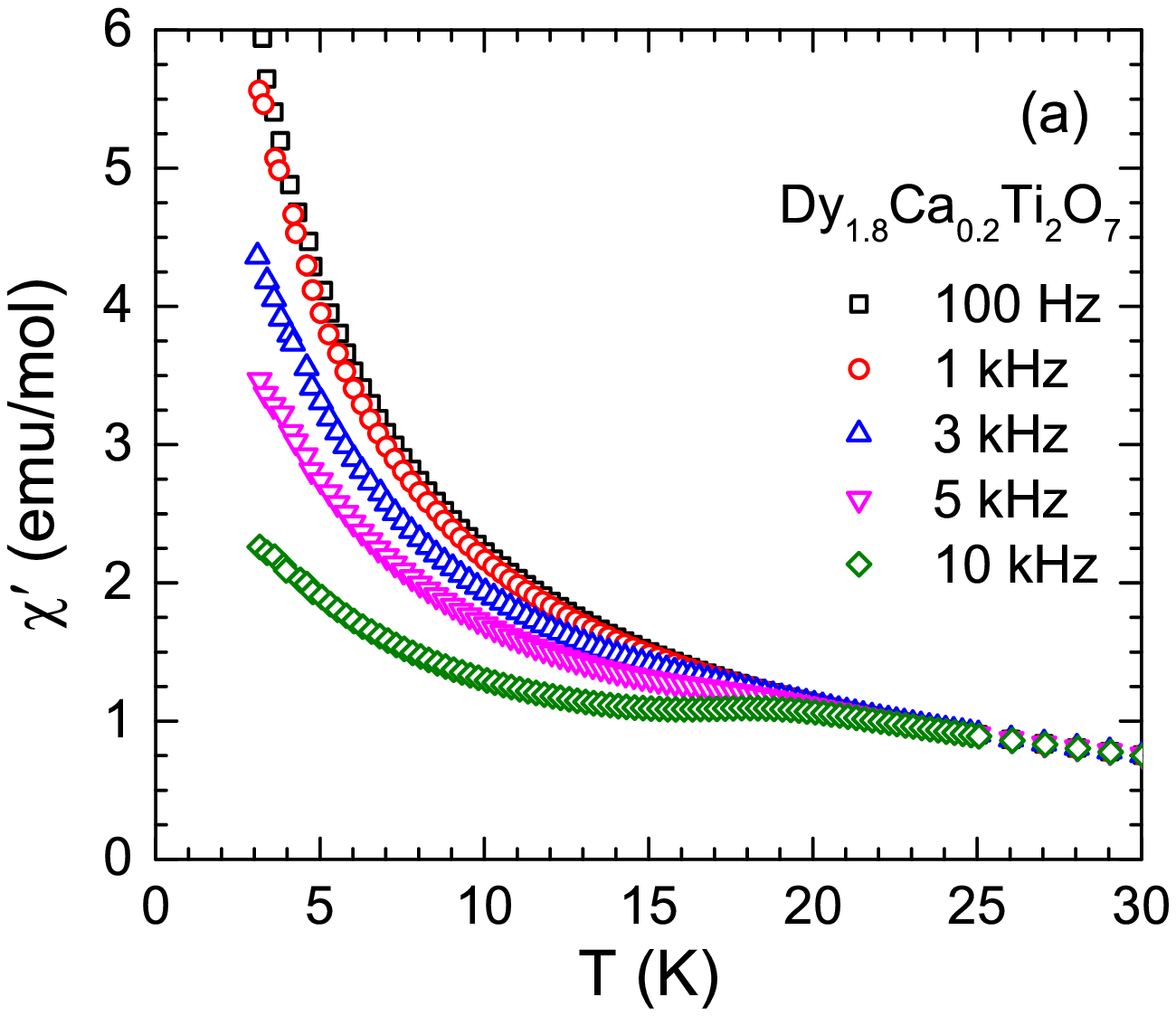}\vspace{0.3cm}
\includegraphics[width=3in, keepaspectratio]{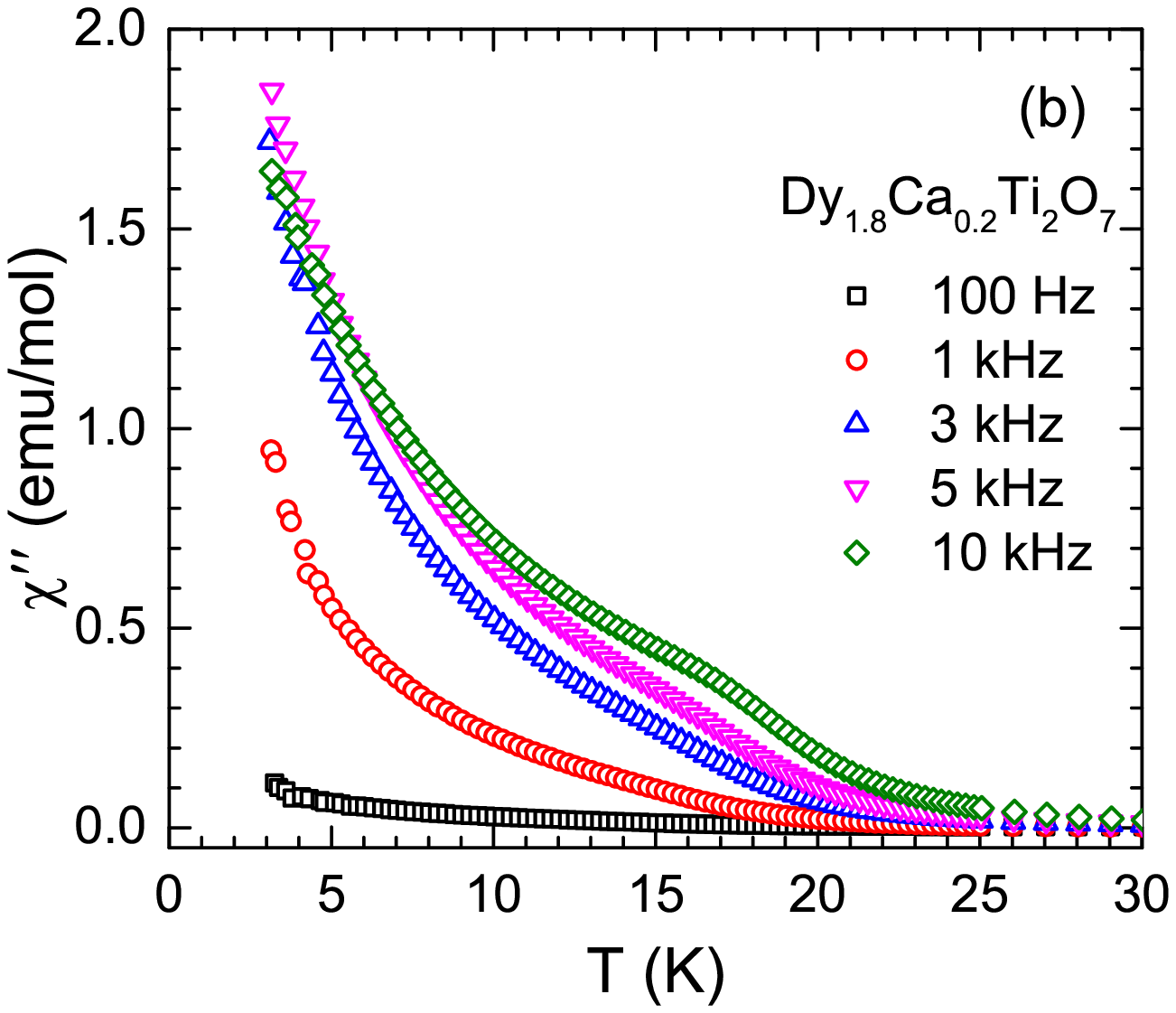}
\caption {(Color online) The temperature $T$ dependence of (a) real $\chi'$ and (b) imaginary $\chi''$ parts of the ac magnetic susceptibility $\chi_{\rm ac}$ of Dy$_{1.8}$Ca$_{0.2}$Ti$_{2}$O$_{7}$ measured at different frequencies from 100~Hz to 10~kHz in an applied ac magnetic field of 1.5~mT\@.}
\label{fig:Chiac1}
\end{center}
\end{figure}

\begin{figure}
\begin{center}
\includegraphics[width=3in, keepaspectratio]{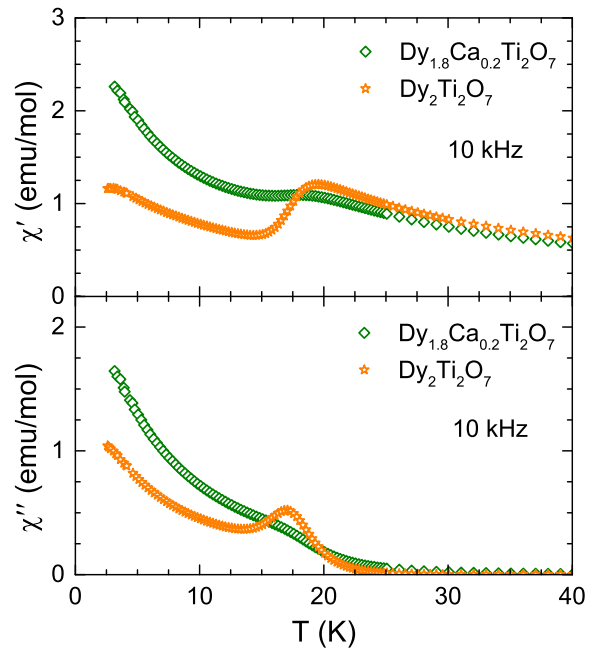}
\caption {(Color online) The temperature $T$ dependence of real $\chi'$ and  imaginary $\chi''$ parts of the ac magnetic susceptibility $\chi_{\rm ac}$ of Dy$_{1.8}$Ca$_{0.2}$Ti$_{2}$O$_{7}$ and Dy$_2$Ti$_2$O$_7$ measured at 10~kHz in an applied ac magnetic field of 1.5~mT\@.}
\label{fig:Chiac2}
\end{center}
\end{figure}

The temperature $T$ dependence of ac magnetic susceptibility $\chi_{\rm ac}$ of Dy$_{1.8}$Ca$_{0.2}$Ti$_{2}$O$_{7}$ measured in an ac field of 1.5~mT at different frequencies between 100~Hz to 10~kHz is shown in figure~\ref{fig:Chiac1}. It is seen from figure~\ref{fig:Chiac1} that both the real part $\chi'$ and imaginary part $\chi''$ of $\chi_{\rm ac}$ show frequency dependent behavior. While at low frequencies there is no anomaly in $\chi'(T)$ or $\chi''(T)$, clear anomalies related to spin freezing are seen at higher frequencies. At 10~kHz an anomaly in $\chi'(T)$ is seen at around 19~K [figure~\ref{fig:Chiac1}(a)] and that in $\chi''(T)$ near 17~K [figure~\ref{fig:Chiac1}(b)]. This kind of frequency dependent behavior of $\chi_{\rm ac}$, i.e. appearance of anomalies in $\chi'(T)$ and $\chi''(T)$ at higher frequencies have also been observed in Dy$_{2}$Ti$_{2}$O$_{7}$ \cite{Snyder2001,Matsuhira2001}. However, the frequency dependence is much weaker in Dy$_{1.8}$Ca$_{0.2}$Ti$_{2}$O$_{7}$ than that in Dy$_{2}$Ti$_{2}$O$_{7}$. While at 1~kHz a well pronounced anomaly can be seen in the case of Dy$_{2}$Ti$_{2}$O$_{7}$ \cite{Snyder2001,Matsuhira2001}, in the case of Dy$_{1.8}$Ca$_{0.2}$Ti$_{2}$O$_{7}$ no detectable anomaly is seen at 1~kHz in $\chi'(T)$ and anomaly in $\chi''(T)$ is very weak. The $\chi'(T)$ and $\chi''(T)$ data of Dy$_{1.8}$Ca$_{0.2}$Ti$_{2}$O$_{7}$ and Dy$_2$Ti$_2$O$_7$ at 10 kHz are compared in figure~\ref{fig:Chiac2}. Even though the anomaly temperatures of $\chi'(T)$ and $\chi''(T)$ are nearly same in both cases, the relative changes in magnitudes at these temperatures are much weaker in Dy$_{1.8}$Ca$_{0.2}$Ti$_{2}$O$_{7}$. The frequency dependence of $\chi_{\rm ac}(T)$  for Dy$_{1.8}$Ca$_{0.2}$Ti$_{2}$O$_{7}$ is substantially weaker than the case of Y substituted Dy$_{2-x}$Y$_{x}$Ti$_{2}$O$_{7}$ \cite{Snyder2002}. Thus Ca$^{2+}$ substitution seems more effective in suppressing the spin-freezing behavior in Dy$_{2}$Ti$_{2}$O$_{7}$ than Y$^{3+}$ substitution.

\begin{figure}
\begin{center}
\includegraphics[width=3in, keepaspectratio]{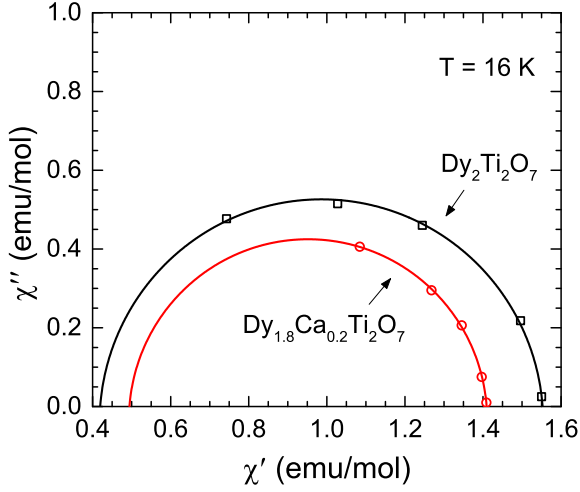}
\caption {(Color online) Cole-Cole plot (imaginary part $\chi''$ versus real part $\chi'$) of the ac magnetic susceptibility for Dy$_{1.8}$Ca$_{0.2}$Ti$_{2}$O$_{7}$ and Dy$_2$Ti$_2$O$_7$ at different frequencies at 16~K\@.}
\label{fig:Chiac3}
\end{center}
\end{figure}

The Cole-Cole plot of $\chi''$ versus $\chi'$ for Dy$_{1.8}$Ca$_{0.2}$Ti$_{2}$O$_{7}$ is shown in figure~\ref{fig:Chiac3} for $T=16$~K\@. The Cole-Cole plot depicts the spin-dynamics at a given temperature and accesses the distribution of relaxation times of spin-freezing processes. While a semicircular character of plot indicates a single relaxation time or a narrow distribution of relaxation times, the semicircle becomes flattened and/or distorted when there is a distribution of relaxation times. From figure~\ref{fig:Chiac3} we see that the Cole-Cole plot for Dy$_{1.8}$Ca$_{0.2}$Ti$_{2}$O$_{7}$ has a semicircular character similar to that of Dy$_{2}$Ti$_{2}$O$_{7}$, though the center of the circle is not same for the two. Within the Cole-Cole formalism, the $\chi''(\chi')$ data can be fitted to the equation of semicircular arc given by \cite{Mydosh1993}
\begin{eqnarray}
\chi''(\chi')& =-\frac{\chi_0 - \chi_{\rm s}}{2 \tan[(1-\alpha)\frac{\pi}{2}]} \nonumber \\
             & + \sqrt{(\chi'-\chi_{\rm s})(\chi_0-\chi')+\frac{(\chi_0 - \chi_{\rm s})^2}{4 \tan^2[(1-\alpha)\frac{\pi}{2}]}}. 
\label{eq:Cole-Cole}
\end{eqnarray}
Here $\chi_0$ is the isothermal susceptibility (in the limit of low frequency) and $\chi_{\rm s}$ the adiabatic susceptibility (in the limit of high frequency).  The parameter $\alpha$ represents the width of the distribution of relaxation times, $\alpha= 0$ implies a single relaxation time and $\alpha= 1$ an infinitely wide distribution.  The fits of $\chi''(\chi')$ by Eq.~(\ref{eq:Cole-Cole}) are shown by solid curves in figure~\ref{fig:Chiac3}. From the fits we obtained $\chi_0 = 1.409(2)$~emu/mol, $\chi_{\rm s} = 0.49(3)$~emu/mol and $(1-\alpha){\pi}/{2} = 89.46(2)^\circ$ for Dy$_{1.8}$Ca$_{0.2}$Ti$_{2}$O$_{7}$,  and $\chi_0 = 1.552(4)$~emu/mol, $\chi_{\rm s} = 0.42(3)$~emu/mol and $(1-\alpha){\pi}/{2} = 89.46(3)^\circ$ for Dy$_{2}$Ti$_{2}$O$_{7}$. Accordingly we obtain $\alpha \approx 0.006$ which is very close to 0 for both samples.  The different values of $\chi_0$ and $\chi_{\rm s}$ reflect different center and radius of the semicircular arc for Dy$_{1.8}$Ca$_{0.2}$Ti$_{2}$O$_{7}$ and Dy$_{2}$Ti$_{2}$O$_{7}$. However, $\alpha$ is identical for both. The semicircular nature of Cole-Cole plot and the extremely small values of $\alpha$ suggest a single relaxation mode in Dy$_{1.8}$Ca$_{0.2}$Ti$_{2}$O$_{7}$ as well as in Dy$_{2}$Ti$_{2}$O$_{7}$ \cite{Snyder2001}.

\begin{figure}
\begin{center}
\includegraphics[width=3in, keepaspectratio]{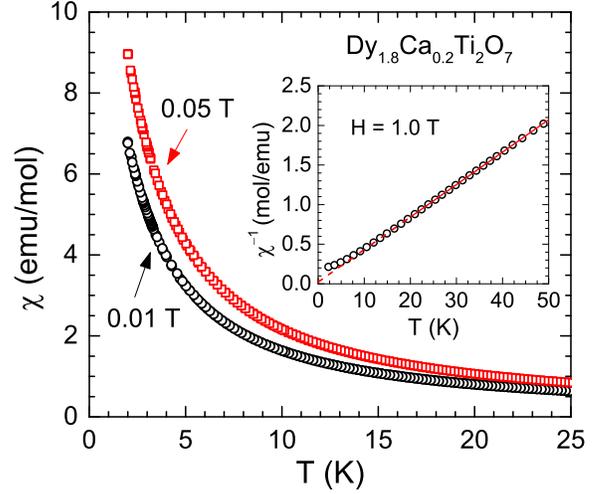}
\caption{(Color online) Zero-field-cooled magnetic susceptibility $\chi$ of Dy$_{1.8}$Ca$_{0.2}$Ti$_{2}$O$_{7}$ as a function of temperature $T$ for $2 \leq T \leq 25$~K measured in magnetic fields $H= 0.01$~T and 0.05~T\@. Inset: Inverse magnetic susceptibility $\chi^{-1}(T)$ for $2 \leq T \leq 50$~K in $ H = 1.0$~T\@. The solid red line is the fit of the $\chi^{-1}(T)$ data by the Curie-Weiss law in $20~{\rm K} \leq T \leq 50$~K and the dashed line is an extrapolation.}
\label{fig:MT}
\end{center}
\end{figure}

The zero-field-cooled dc magnetic susceptibility $\chi(T)$ of Dy$_{1.8}$Ca$_{0.2}$Ti$_{2}$O$_{7}$ measured in $H = 0.01$~T and 0.05~T are shown in figure~\ref{fig:MT}. The $\chi(T)$ data do not show any anomaly near $T_f$ and paramagnetic behavior is evident from the $\chi(T)$ data at $T \geq 2$~K. The magnitude of $\chi$ at low-$T$ is quite large. A similar $\chi(T)$ behavior has been observed for Dy$_2$Ti$_2$O$_7$ \cite{Bramwell2000}. From figure~\ref{fig:MT} it is evident that the $\chi(T)$ measured at 0.05~T is higher in magnitude than that at 0.01~T\@. This reflects the ferromagnetic nature of magnetic exchange as in the Dy$_2$Ti$_2$O$_7$. At high temperature, the $\chi(T)$ data of Dy$_{1.8}$Ca$_{0.2}$Ti$_{2}$O$_{7}$ follow Curie-Weiss behavior, $\chi(T) = C/(T-\theta_{\rm p})$. A linear fit of inverse magnetic susceptibility $\chi^{-1}(T)$ data (measured in $ H = 1.0$~T) over 150~K~$\leq T\leq$~300~K yields the Curie constant $C = 28.22(5)$~emu\,K/mol and Weiss temperature $\theta_{\rm p} = -14.9(6)$~K. The $\chi^{-1}(T)$ data when fitted over 20~K~$\leq T\leq$~50~K yield $\theta_{\rm p} = -0.55(1)$~K and $C = 24.38(1)$~emu\,K/mol. The linear fit of the $H=1.0$~T $\chi^{-1}(T)$ data in 20~K~$\leq T\leq$~50~K is shown by solid red line in the inset of figure~\ref{fig:MT}.  Like Dy$_{1.8}$Ca$_{0.2}$Ti$_{2}$O$_{7}$ two different values of $\theta_{\rm p}$ have also been found in the case of Dy$_2$Ti$_2$O$_7$: $\theta_{\rm p}= -15$~K and $-0.8$~K from high-$T$ and low-$T$ fit ranges, respectively \cite{Jana2002}. A positive $\theta_{\rm p}$ has been reported for Dy$_2$Ti$_2$O$_7$ from the analysis of $\chi(T)$ data below 40~K \cite{Ramirez1999,Fukazawa2002}. Demagnetizing effect is argued to cause the $\theta_{\rm p}$ to be negative and when corrected for this the $\theta_{\rm p}$ would be positive such that effectively there is a weak ferromagnetic interaction in Dy$_2$Ti$_2$O$_7$ \cite{Bramwell2000}. A similar situation should be applicable to present data. Our sample for dc magnetic susceptibility measurements consisted of several small pieces of arbitrary shape and sizes therefore we do not have a good estimate of demagnetization factor to correct the $\chi(T)$ data for the demagnetizing field. A crude approximation assuming spherical shape with demagnetization factor $N = 1/3$ yields $\theta_{\rm p} = 0.73(1)$~K on fitting the corrected $\chi(T)$ data by Curie-Weiss law in 20~K~$ \leq T \leq 50$~K with $C = 24.39(1)$~emu\,K/mol. Thus the demagnetization correction gives a positive $\theta_{\rm p}$ and hence a ferromagnetic exchange. We estimate an effective moment $\mu_{\rm eff} = 10.41\, \mu_{\rm B}$/Dy from the value of $C = 24.38$~emu\,K/mol which is very close to the theoretically expected value of $10.61 \, \mu_{\rm B}$ for $^6$G$_{15/2}$ ground state of Dy$^{3+}$ ions.

\begin{figure}
\begin{center}
\includegraphics[width=3in, keepaspectratio]{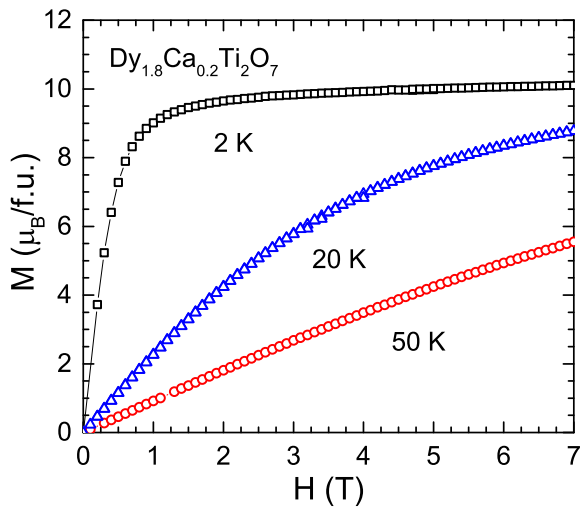}
\caption{(Color online) Isothermal magnetization $M$ of Dy$_{1.8}$Ca$_{0.2}$Ti$_{2}$O$_{7}$ as a function of applied magnetic field $H$ for $0 \leq H \leq 7$~T measured at the indicated temperatures.}
\label{fig:MH}
\end{center}
\end{figure}

The isothermal $M(H)$ data of Dy$_{1.8}$Ca$_{0.2}$Ti$_{2}$O$_{7}$ measured at selected temperatures of 2~K, 20~K, and 50~K are shown in figure~\ref{fig:MH}. At 2~K the $M(H)$ isotherm exhibits a rapid increase below 0.5~T and tends to saturate thereafter with a saturation value of magnetization $M_s = 5.61\, \mu_{\rm B}$/Dy at 7~T\@. This value of $M_s $ is close to the value expected for [100] direction $M_s = 5.77\, \mu_{\rm B}$/Dy on account of FM interaction and Ising anisotropy \cite{Fukazawa2002}. However, the $M_s $ value is much lower than the theoretical value of $M_{\rm sat} = g_J J\,\mu_{\rm B} = 10\,\mu_{\rm B}$ for free Dy$^{3+}$ ions ($g_J = 4/3$ and $J = 15/2$), in line with the presence of strong Ising anisotropy in this compound. The saturation behavior of $M(H)$ isotherms depends on the temperature: $M_s $ decreases with increasing temperature. The overall $M(H)$ behavior of Dy$_{1.8}$Ca$_{0.2}$Ti$_{2}$O$_{7}$ is similar to that of Dy$_2$Ti$_2$O$_7$ \cite{Bramwell2000}. For a polycrystalline Dy$_2$Ti$_2$O$_7$, $M_s \approx 5\,\mu_{\rm B}$/Dy at 1.8~K \cite{Snyder2004}. Thus we see an increase of about 12\% in $M_s $ value at 2~K with 10\% Ca substitution for Dy. This effect of Ca substitution is quite distinct from Y substitution where dilution has negligible effect on saturation moment [15]. 

In Dy$_2$Ti$_2$O$_7$ there is a strong Ising anisotropy along the local $\langle 111 \rangle$ direction that determines the ice rule and hence the magnetic behavior of this compound. The $M(H)$ data of Dy$_2$Ti$_2$O$_7$ obey ice rule (two-in, two-out configuration) below 0.5~T and at fields higher than this Zeeman energy exceeds the effective nearest neighbor ferromagnetic interaction which causes a breaking of ice rule along the [111] direction \cite{Fukazawa2002}. The [111] direction  $M(H)$ data for $H > 2.5$~T with $M_s \approx 5\,\mu_{\rm B}$/Dy at 1.8 K (much higher than the ice rule expected value of 3.33\,$\mu_{\rm B}$/Dy) reveal three-in, one-out spin configuration, clearly reflecting the breaking of ice rule at high fields \cite{Fukazawa2002}. For a spin ice system with two-in, two-out configuration the predicted values of $M_s $ for the [100], [110] and [111] directions are $M_{\rm sat}/\surd 3$ ($= 5.77\,\mu_{\rm B}$/Dy), $M_{\rm sat}/\surd 6$ ($= 4.08\,\mu_{\rm B}$/Dy) and $M_{\rm sat}/3$ ($=3.33\,\mu_{\rm B}$/Dy), respectively, where $M_{\rm sat}$ is the full saturation moment \cite{Fukazawa2002}. Therefore for two-in, two-out configuration the weighted average $M_s $ value for the polycrystalline sample should be around $4.24\,\mu_{\rm B}$/Dy according to $\langle M_s\rangle = (6\,M_{[100]} + 12\, M_{[110]} + 8\, M_{[111]})/26$, where the weight factors 6, 12 and 8 are the number of equivalent directions for [100], [110] and [111], respectively. Our value of $M$ for Dy$_{1.8}$Ca$_{0.2}$Ti$_{2}$O$_{7}$ is $4.04\,\mu_{\rm B}$/Dy at 0.5~T and 2 K which is compatible with the theoretically expected ice rule value [$4.24\,\mu_{\rm B}$/Dy] and is very close to the measured weighted average value (over [100], [110] and [111] directions) of  $M \approx 4.0\,\mu_{\rm B}$/Dy for single crystal Dy$_2$Ti$_2$O$_7$ at 0.5~T and 1.8~K \cite{Fukazawa2002}. However, the observed $M$ at 0.5~T is about 14\% higher than the $M \approx 3.5\,\mu_{\rm B}$/Dy reported for polycrystalline Dy$_2$Ti$_2$O$_7$ at 0.5~T and 1.8~K in Ref.~\cite{Snyder2004}. Thus, the saturation behavior of $M(H)$ of Dy$_{1.8}$Ca$_{0.2}$Ti$_{2}$O$_{7}$ is substantially different from both undoped as well as Y-doped Dy$_2$Ti$_2$O$_7$ \cite{Snyder2004}. While we do not have a clear explanation for the increased $M_s$ for Dy$_{1.8}$Ca$_{0.2}$Ti$_{2}$O$_{7}$ we suspect that a change in crystal field effect or a change in exchange interaction by Ca substitution is the origin for the observed behavior.

The charge/valence consideration would suggest three possible situations: 1) (${\rm Dy^{3+})_{1.8}(Ca^{2+})_{0.2}(Ti^{4+})_2(O^{2-})_7}$ i.e.\ with a deficiency of positive charge resulting in a hole doping. 2) (${\rm Dy^{3+})_{1.8}(Ca^{2+})_{0.2}(Ti^{4+})_2(O^{2-})_{7-\delta}}$ i.e.\ a deficiency of oxygen concentration to balance the charge such that the composition is Dy$_{1.8}$Ca$_{0.2}$Ti$_{2}$O$_{6.9}$ instead of Dy$_{1.8}$Ca$_{0.2}$Ti$_{2}$O$_{7}$. 3) (${\rm Dy^{3+})_{1.8}(Ca^{2+})_{0.2}(Ti^{(4+\epsilon)+})_2(O^{2-})_7}$  i.e.\ a change in valence state of Ti, however this would require an increase in valence of Ti to compensate for loss in positive charge at Dy/Ca site if there is no change in oxygen concentration. We would like to point out that Ti is not known to have a valence state higher than 4+, therefore this seems very unlikely. On the other hand from the view point of charge (valence) balancing an oxygen deficiency is possible. However, an oxygen deficiency in Dy$_2$Ti$_2$O$_7$ has been found to reduce the saturation magnetization \cite{Sala2014}. Contrary to such an expectation, an increase in saturation magnetization in Dy$_{1.8}$Ca$_{0.2}$Ti$_{2}$O$_{7}$  does not support the oxygen deficiency in this compound. An increase in saturation magnetization can be brought by oxygen deficiency if the deficiency concentration is more than what is required to balance the loss in positive charge upon Ca substitution such that a fraction of Ti becomes magnetic, i.e.\ Ti$^{3+}$ and the increased magnetization comes from Ti. Another possible origin of excess magnetization can be canting of Dy$^{3+}$ moments from the easy $\langle 111 \rangle$ direction upon the partial removal of Dy$^{3+}$ spins from the vertices of a tetrahedron with two-in, two-out spin configuration as a result of Ca$^{2+}$ substitution. Ca$^{2+}$ is larger than Dy$^{3+}$ in size and may distort the crystal field, changing slightly the Dy$^{3+}$ wavefunction and resulting anisotropy. If one of four Dy is replaced by Ca then such tetrahedra will have either two-in, one-out or one-in, two-out configuration. Thus assuming uniform distribution of Ca, about 10\% of the tetrahedra in Dy$_{1.8}$Ca$_{0.2}$Ti$_{2}$O$_{7}$ will no longer fulfil the two-in, two-out spin configuration and when these tetrahedra are subjected to high magnetic fields they may behave differently, canting of spins could be one such situation.

\begin{figure}
\begin{center}
\includegraphics[width=3in, keepaspectratio]{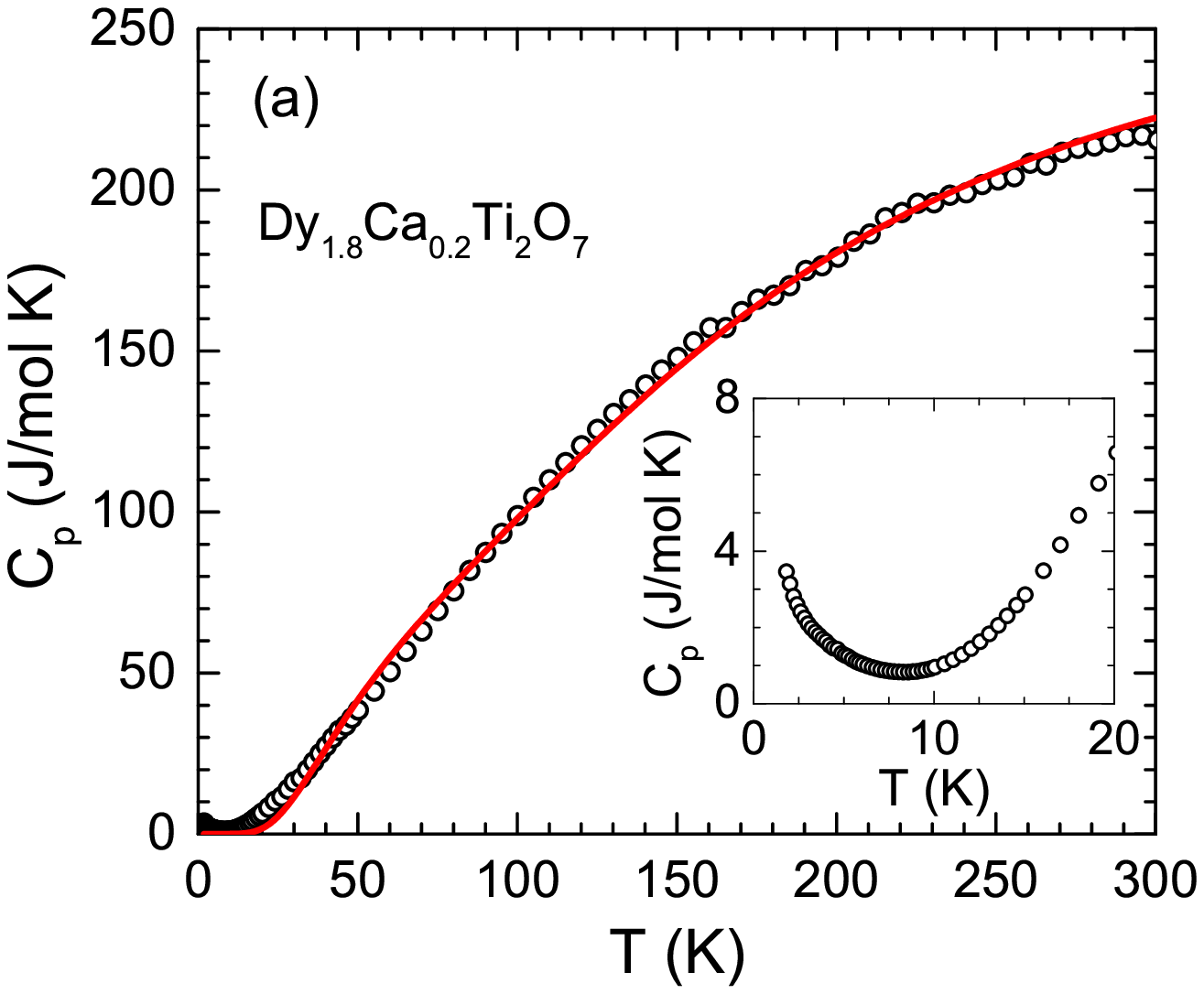}\vspace{0.3cm}
\includegraphics[width=3in, keepaspectratio]{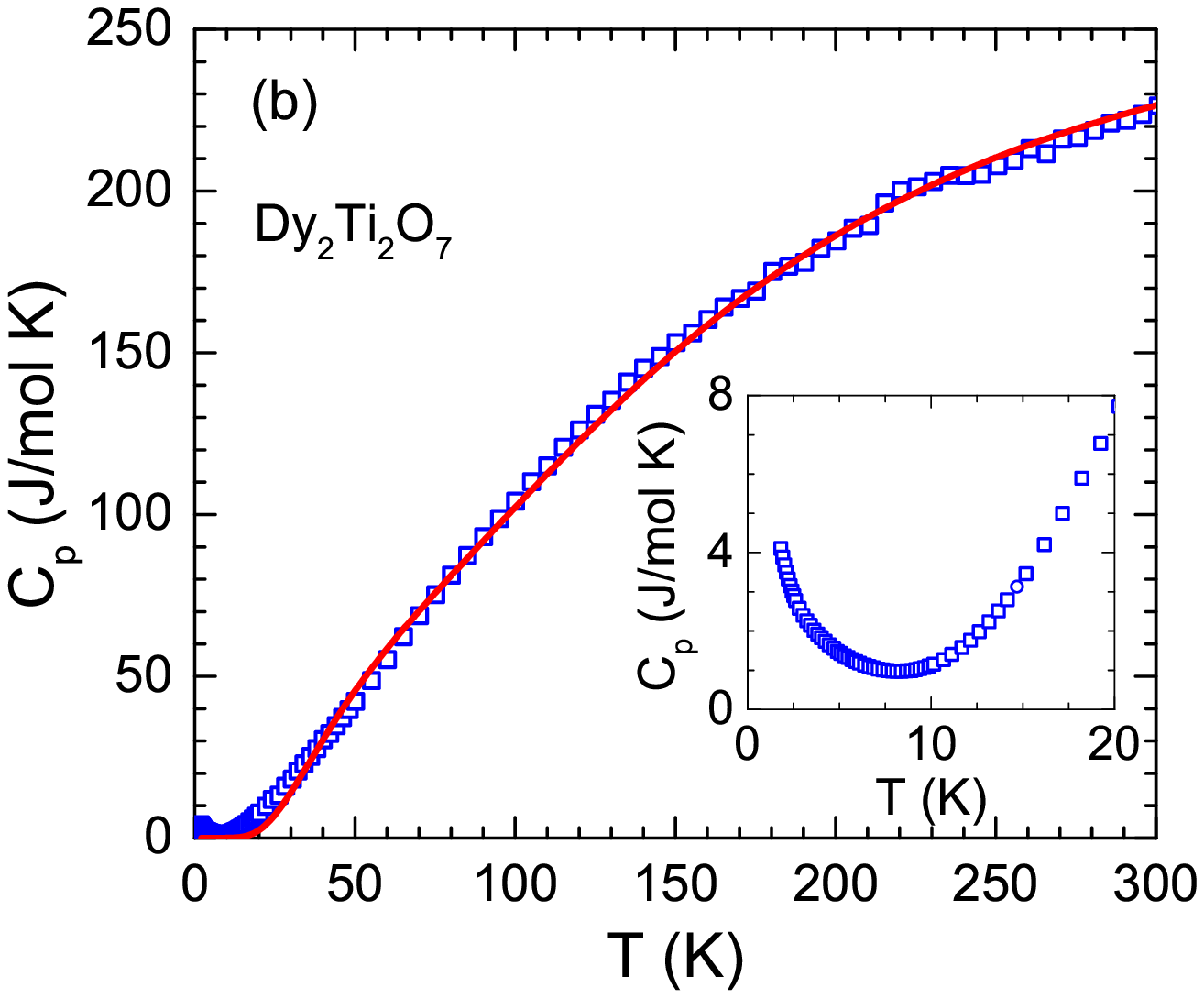}\vspace{0.3cm}
\includegraphics[width=3in, keepaspectratio]{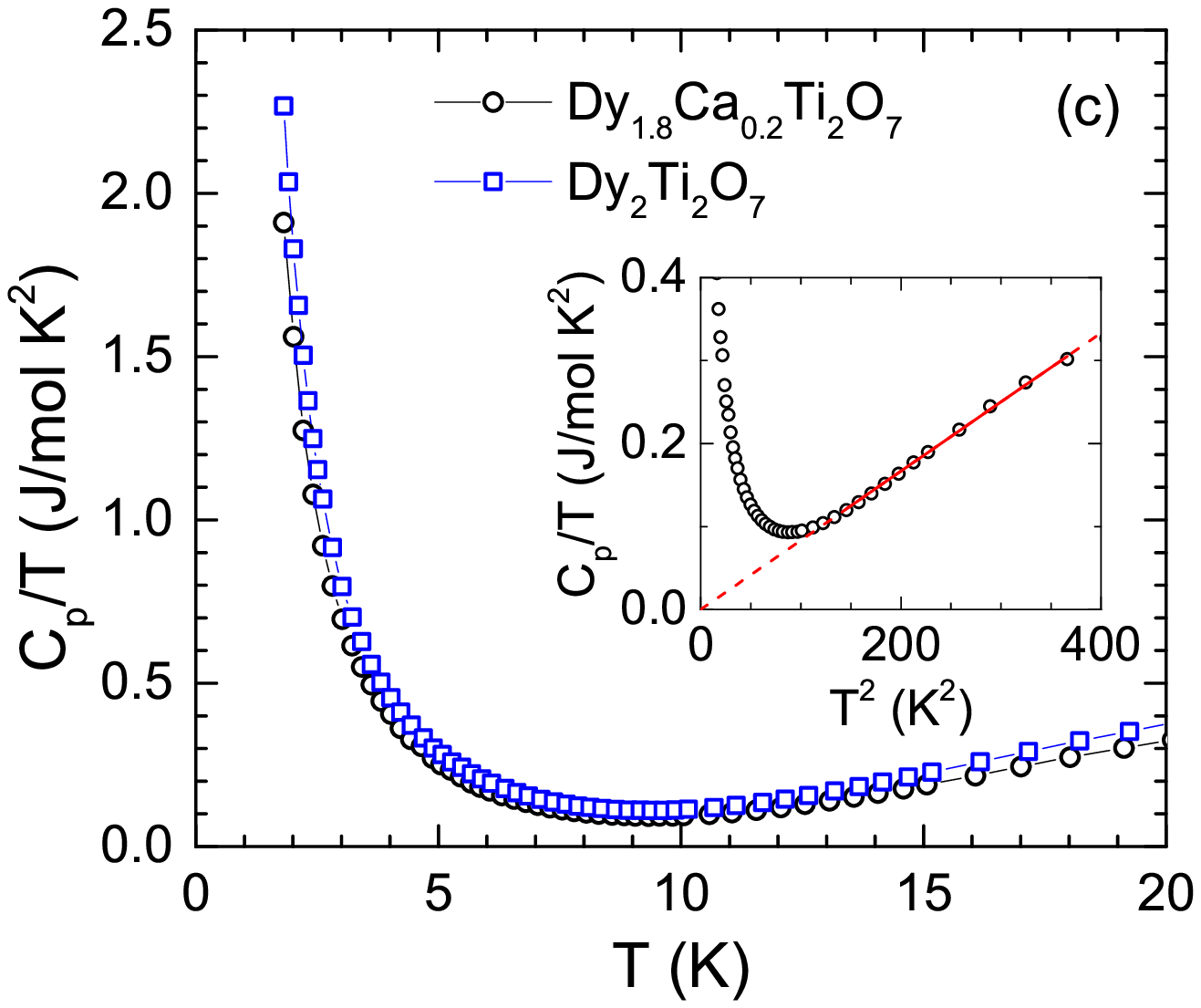}
\caption{\label{fig:HC} (Color online) Heat capacity $C_{\rm p}$ as a function of temperature $T$ for 1.8~K~$\leq T \leq$~300~K measured in zero field for (a) Dy$_{1.8}$Ca$_{0.2}$Ti$_{2}$O$_{7}$ and (b) Dy$_{2}$Ti$_{2}$O$_{7}$. The red solid curves are the sum of the contributions from the Debye+Einstein models and the crystal electric field according to Eq.~\ref{eq:HC-fit}. Insets: Low-$T$ $C_{\rm p}(T)$ for $T \leq 20.0$~K\@. (c) $C_{\rm p}/T$ versus $T$ plot of \mbox{low-$T$} $C_{\rm p}(T)$ data of Dy$_{1.8}$Ca$_{0.2}$Ti$_{2}$O$_{7}$ and Dy$_{2}$Ti$_{2}$O$_{7}$ in $1.8~{\rm K} \leq T \leq 20$~K\@. Inset: $C_{\rm p}/T$ versus $T^2$ plot for Dy$_{1.8}$Ca$_{0.2}$Ti$_{2}$O$_{7}$ for $T \leq 20$~K on an expanded $y$-scale. The solid red line is the fit to $C_{\rm p}/T = \beta T^2$  in $12~{\rm K} \leq T \leq 19$~K and the dashed lines are extrapolations.}
\end{center}
\end{figure}

The $C_{\rm p}(T)$ data of Dy$_{1.8}$Ca$_{0.2}$Ti$_{2}$O$_{7}$ are shown in figure~\ref{fig:HC}(a). It is seen from figure~\ref{fig:HC}(a) that the $C_{\rm p}(T)$ data at $T = 300$~K is much lower than the classical Dulong-Petit value $C_{\rm V} = 3nR = 33R \approx 274.4$~J/mol\,K, where $R$ is the molar gas constant and $n=11$ is the number of atoms per formula unit \cite{Kittel2005, Gopal1966}. The $C_{\rm p}(T)$ data of Dy$_{2}$Ti$_{2}$O$_{7}$ shown in figure~\ref{fig:HC}(b) also fails to reach the high-$T$ Dulong-Petit limit by 300~K\@. The lower value of $C_{\rm p}$ at $T = 300$~K than its high-$T$ limit is consistent with the high Debye temperature ($\Theta_{\rm D}$ much greater than 300~K) of pyrochlore 227 compounds. Like $\chi(T)$, the $C_{\rm p}(T)$ data do not show any anomaly near $T_f$.  At low-$T$ an increase is observed in the $C_{\rm p}$ of Dy$_{1.8}$Ca$_{0.2}$Ti$_{2}$O$_{7}$ with decreasing $T$ at $T \leq 8$~K [inset of figure~\ref{fig:HC}(a)]. This behavior of $C_{\rm p}(T)$ is similar to that of Dy$_{2}$Ti$_{2}$O$_{7}$ [inset of figure~\ref{fig:HC}(b)]. A comparison of $C_{\rm p}/T$ versus $T$ plot for Dy$_{1.8}$Ca$_{0.2}$Ti$_{2}$O$_{7}$ and Dy$_{2}$Ti$_{2}$O$_{7}$ in figure~\ref{fig:HC}(c) shows that there is almost no change in upturn feature of $C_{\rm p}(T)$ by 10\% Ca substitution.

At low-$T$ heat capacity is usually expressed by $C_{\rm p}(T) = \gamma T +\beta T^{3}$, where $\gamma$ and $\beta$ are coefficients to the electronic and lattice contributions to heat capacity, respectively. The Sommerfeld coefficient $\gamma$ is nonzero for a metallic sample. However, for an insulating material $\gamma = 0$, therefore $C_{\rm p}(T)$ can be described by $\beta T^{3}$ term only. We fitted the $C_{\rm p}(T)$ data of Dy$_{1.8}$Ca$_{0.2}$Ti$_{2}$O$_{7}$ by $C_{\rm p}(T) = \gamma T +\beta T^{3}$ and found that $\gamma$ is indeed zero which is consistent with the insulating characteristic of  Dy$_{1.8}$Ca$_{0.2}$Ti$_{2}$O$_{7}$. The fit of $C_{\rm p}/T$ versus $T^2$ by  $C_{\rm p}/T = \beta T^2$  in  $12~{\rm K} \leq T \leq 19$~K is shown by solid red line in the inset of figure~\ref{fig:HC}(c) which gives $\beta=8.34(3) \times 10^{-4}$~J/mol\,K$^{4}$. A similar analysis of $C_{\rm p}(T)$ of Dy$_2$Ti$_2$O$_7$ gives $\beta=9.78(5) \times 10^{-4}$~J/mol\,K$^{4}$ in very good agreement with the reported value of $\beta=9.7 \times 10^{-4}$~J/mol\,K$^{4}$ \cite{Hiroi2003}. An estimate of Debye temperature $\Theta_{\rm D}$ follows from $\beta$, as $\Theta_{\rm D} = (12 \pi^{4} n R /{5 \beta} )^{1/3}$, accordingly from the low-$T$ $C_{\rm p}(T)$ data we get $\Theta_{\rm D} \approx 280$~K for Dy$_2$Ti$_2$O$_7$ and $\Theta_{\rm D} \approx 295$~K for Dy$_{1.8}$Ca$_{0.2}$Ti$_{2}$O$_{7}$. The value of $\Theta_{\rm D}$ for Dy$_2$Ti$_2$O$_7$ is in very good agreement with the reported value of $\Theta_{\rm D} = 283$~K obtained from the $C_{\rm p}(T)$ data below 30~K  \cite{Klemke2011}.

The $\Theta_{\rm D}$ values obtained above failed to fit the high-$T$ $C_{\rm p}(T)$ data using the Debye model lattice heat capacity due to acoustic phonons therefore we analyzed the $C_{\rm p}(T)$ data using a combination of the Debye model and Einstein model of phonon heat capacity, thus accounting for both acoustic and optic modes of lattice vibration. The $C_{\rm p}(T)$ data  were fitted by
\begin{equation}
C_{\rm p}(T) = m C_{\rm V\,Debye}(T) + (1-m)C_{\rm V\,Einstein}(T) + C_{\rm CEF}(T)
\label{eq:HC-fit}
\end{equation}
where $C_{\rm V\, Debye}(T)$ represents the Debye lattice heat capacity, $C_{\rm V\, Einstein}(T)$ represents the Einstein lattice heat capacity and $C_{\rm CEF}(T)$ represents the crystal electric field contribution to heat capacity. The phonon contributions $C_{\rm V\, Debye}(T)$ and $C_{\rm V\, Einstein}(T)$ are given by \cite{Gopal1966}
\begin{equation}
C_{\rm{V\,Debye}}(T)=9 n R \left( \frac{T}{\Theta_{\rm{D}}} \right)^3 {\int_0^{\Theta_{\rm{D}}/T} \frac{x^4 e^x}{(e^x-1)^2}\,dx},
\label{eq:Debye_HC}
\end{equation}
and
\begin{equation}
C_{\rm{V\,Einstein}}(T)=3 n R \left( \frac{\Theta_{\rm{E}}}{T} \right)^2  \frac{e^{\Theta_{\rm E}/T}}{(e^{\Theta_{\rm E}/T}-1)^2},
\label{eq:Eistein_HC}
\end{equation}
where $\Theta_{\rm E}$ is the Einstein temperature. For a two-level system $C_{\rm CEF}(T)$ is given by \cite{Gopal1966}
\begin{equation}
C_{\rm CEF}(T)  = R \left(\frac{\Delta}{T}\right)^2 \frac{g_0g_1 {\rm e}^{-\Delta/T}}{[g_0 + g_1 {\rm e}^{-\Delta/T}]^2},
\label{eq:HC-CEF_TwoLevel}
\end{equation}
where $g_0$ is the degeneracy of the ground state, $g_1$ is the degeneracy of first excited state, and $\Delta$ is CEF splitting energy between the ground state and the first excited state.

Under the action of CEF the ($2J+1$)-fold degenerate ground state of Dy$^{3+}$ ($J = 15/2$) splits into eight doublets, therefore $g_0 = g_1 = 2$. There are few estimates of CEF splitting energy $\Delta$ for Dy$_2$Ti$_2$O$_7$, however the CEF parameters have not been determined precisely by inelastic neutron scattering. The reported values of $\Delta$ are different ranging from 100~cm$^{-1}$ to 287~cm$^{-1}$ \cite{Jana2002,Rosenkranz2000,Malkin2004,Lummen2008}. We use the value $\Delta =100$~cm$^{-1}$ determined from the CEF analysis of magnetic susceptibility data \cite{Jana2002}. The CEF contribution is only a few percent of the total $C_{\rm p}$ therefore uncertainty in estimate of $C_{\rm CEF}(T)$ is not expected to have significant effect on the analysis of the phonon contribution using Debye+Einstein models.

The $C_{\rm CEF}(T)$ was calculated using Eq.~\ref{eq:HC-CEF_TwoLevel} for $g_0 = g_1 = 2$ and $\Delta =100$~cm$^{-1}$ and subtracted off from the measured data and then the [$C_{\rm p}(T) - C_{\rm CEF}(T)$] data of Dy$_2$Ti$_2$O$_7$ were fitted by Debye+Einstein models according to Eqs.~(\ref{eq:HC-fit})--(\ref{eq:Eistein_HC}). A reasonable fit is obtained for $m=0.71(1)$ i.e.\ with 71\% weight to Debye term and 29\% to Einstein term. The sum of $C_{\rm CEF}(T)$ and Debye+Einstein models fit for Dy$_2$Ti$_2$O$_7$ is shown by solid red curve in figure~\ref{fig:HC}(b). While fitting the data we used analytic Pad\'e approximant for $C_{\rm V\,Debye}(T)$ as developed in Ref.~\cite{Goetsch2012}. The $\Theta_{\rm D} = 722(8)$~K and $\Theta_{\rm E} = 157(4)$~K were obtained from the fit. For Y$_2$Ti$_2$O$_7$ a value of $\Theta_{\rm D} = 967$~K has been estimated \cite{Johnson2009} using the information from the first principles calculation of elastic properties of Y$_2$Ti$_2$O$_7$, which when scaled according to
\begin{equation}
\Theta_{{\rm D}\,A} = \left( \frac{M_B}{M_A}\right)^{1/2} \left( \frac{V_B}{V_A}\right)^{1/3} \Theta_{{\rm D}\,B}
\label{eq:Debye_T}
\end{equation}
gives $\Theta_{\rm D} = 820$~K for Dy$_2$Ti$_2$O$_7$, where $M$ and $V$ are the mass and volume per formula unit, respectively and the subscripts $A$ and $B$ label the materials $A$ and $B$, respectively. Thus the value of $\Theta_{\rm D} = 722(8)$~K determined from Debye+Einstein models fit for Dy$_2$Ti$_2$O$_7$ is consistent with the expected large value of scaled $\Theta_{\rm D}$ compared to the value of $\Theta_{\rm D}$ determined from coefficient $\beta$ above.

Since not much effect is seen on the $C_{\rm p}(T)$ data by 10\% Ca substitution for Dy, to analyse the $C_{\rm p}(T)$ data of Dy$_{1.8}$Ca$_{0.2}$Ti$_{2}$O$_{7}$ by Debye+Einstein model we used the same $C_{\rm CEF}(T)$ that we estimated for Dy$_2$Ti$_2$O$_7$. The sum of $C_{\rm CEF}(T)$ and Debye+Einstein models fit for Dy$_{1.8}$Ca$_{0.2}$Ti$_{2}$O$_{7}$ is shown by the solid red curve in figure~\ref{fig:HC}(a). In this case we obtained $m=0.69(1)$, however the $\Theta_{\rm D} = 766(9)$~K and $\Theta_{\rm E} = 172(4)$~K obtained from the fit are slightly different from those of Dy$_2$Ti$_2$O$_7$. The increase in $\Theta_{\rm D}$ and $\Theta_{\rm E} $ with Ca substitution is consistent with their formula mass dependence according to Eq.~(\ref{eq:Debye_T}) as the formula mass is lowered by Ca substitution.

\section{\label{Conclusion} Summary and Conclusions}

The effect of partial nonmagnetic Ca$^{2+}$ substitution for magnetic Dy$^{3+}$ in spin-ice system Dy$_2$Ti$_2$O$_7$ has been investigated by $\chi_{\rm ac}(T)$, $\chi(T)$, $M(H)$ and $C_{\rm p}(T)$ measurements on Dy$_{1.8}$Ca$_{0.2}$Ti$_{2}$O$_{7}$. The high-frequency $\chi_{\rm ac}(T)$ data of Dy$_{1.8}$Ca$_{0.2}$Ti$_{2}$O$_{7}$ show evidence for spin-freezing type anomaly  near 16~K, however no corresponding anomaly is observed in $\chi(T)$ and $C_{\rm p}(T)$ data. The $\chi(T)$ and $C_{\rm p}(T)$ data of Dy$_2$Ti$_2$O$_7$ also do not show any anomaly near 16~K. The low-$T$ $C_{\rm p}(T)$ data give Sommerfeld coefficient $\gamma = 0$ which reflects an insulating ground state for Dy$_{1.8}$Ca$_{0.2}$Ti$_{2}$O$_{7}$ similar to Dy$_2$Ti$_2$O$_7$. Thus the holes produced by 10\% Ca$^{2+}$ substitution for Dy$^{3+}$ ions is not sufficient to change the electrical properties of insulating Dy$_2$Ti$_2$O$_7$. The high-$T$ $C_{\rm p}(T)$ data were analyzed by the Debye+Einstein models of lattice heat capacity along with a CEF contribution which yielded $\Theta_{\rm D} = 766(9)$~K and $\Theta_{\rm E} = 172(4)$~K for Dy$_{1.8}$Ca$_{0.2}$Ti$_{2}$O$_{7}$  and $\Theta_{\rm D} = 722(8)$~K and $\Theta_{\rm E} = 157(4)$~K for Dy$_2$Ti$_2$O$_7$.

The spin-glass type freezing behavior near 16~K in Dy$_2$Ti$_2$O$_7$ is strongly influenced by Ca substitution. The frequency dependent $\chi_{\rm ac}(T)$ data of Dy$_{1.8}$Ca$_{0.2}$Ti$_{2}$O$_{7}$ show that the suppression of spin-freezing behavior by Ca$^{2+}$ is substantially increased compared to that by the nonmagnetic isovalent Y$^{3+}$ in Dy$_{1.8}$Y$_{0.2}$Ti$_{2}$O$_{7}$ \cite{Snyder2002}. An analysis of $\chi_{\rm ac}(T)$ within the Cole-Cole formalism suggests that no change in semicircular nature of Cole-Cole plot of  $\chi''(\chi')$ results from Ca substitution and the spin dynamics of Dy$_{1.8}$Ca$_{0.2}$Ti$_{2}$O$_{7}$ is governed by a single relaxation time like Dy$_{2}$Ti$_{2}$O$_{7}$ \cite{Snyder2001}.

The Ca$^{2+}$ substitution for Dy from the viewpoint of charge (valence) balancing can result in a hole doping or an oxygen deficiency. Since an oxygen deficiency in Dy$_2$Ti$_2$O$_7$ has been found to reduce the saturation magnetization \cite{Sala2014}, the increased saturation magnetization in Dy$_{1.8}$Ca$_{0.2}$Ti$_{2}$O$_{7}$ is counter-intuitive for oxygen deficiency in this compound. As such the hole doping scenario seems more appropriate, however the holes produced by 10\% Ca doping apparently fails to bring any noticeable change in band structure to reveal metallic nonzero $\gamma$ in heat capacity measurements. It should be noted though that we cannot completely rule out the possibility of O deficiency and the compound may have a composition Dy$_{1.8}$Ca$_{0.2}$Ti$_{2}$O$_{7-\delta}$ instead of Dy$_{1.8}$Ca$_{0.2}$Ti$_{2}$O$_{7}$.

In pure Dy$_2$Ti$_2$O$_7$ the spin relaxation has been suggested to exhibit an unusual double crossover, the spin relaxation is thermally activated at temperatures above a crossover temperature $T_{\rm cross} \sim 13$~K below which the relaxation process is dominated by quantum tunneling down to $T_{\rm ice} \sim 4$~K and then again a thermally activated behavior dominates below $T_{\rm ice}$ as the spins become strongly correlated below this temperature \cite{Snyder2003}. Ehlers {\it et al}.\ \cite{Ehlers2003} made an attempt to explain the difference between the 16 K freezing in Dy$_2$Ti$_2$O$_7$ (presence) and Ho$_2$Ti$_2$O$_7$ (absence) based on the competition between the thermally activated and quantum spin tunneling in these compounds. Considering that the spin dynamics near 16 K can be characterized by an Arrhenius type behavior with an activation energy $E_a \sim 220$ K which is close to the CEF splitting energy (240--340 K), the spin freezing is considered to involve transitions to and from the first excited doublet \cite{Gardner2010,Ehlers2003}. Since Dy$^{3+}$ ($J = 15/2$) is a Kramers ion, a direct relaxation process is not allowed unless the degeneracy of ground state doublet is removed (e.g.\ by the application of magnetic field). Therefore an Orbach process is considered as an alternative mechanism of relaxation in Dy$_2$Ti$_2$O$_7$ which is a two stage process in which a spin first absorbs a phonon of the same magnitude as the energy of the excited crystal field state and then relaxes to a state lower in energy than its initial state by emitting a phonon of different energy than the absorbed one. Ho$^{3+}$ ($J = 8$) on the other hand being non-Kramers ion, no Orbach-like relaxation is observed near 16 K in Ho$_2$Ti$_2$O$_7$. Ehlers {\it et al}.\ suggested that Ho$_2$Ti$_2$O$_7$ can reveal 16~K feature under the application of magnetic field as confirmed by the neutron spin echo experiments \cite{Ehlers2003,Ehlers2004}.

Considering the effect of Ca substitution on spin-freezing within the Orbach-like relaxation formalism that involves both the lattice degrees of freedom and the excited crystal field levels for spin-lattice relaxation, it seems that a significant change in the crystal field scheme/energy is brought by the Ca substitution. With a coordination number of 8, the Y$^{3+}$ (101.9 pm) has similar ionic radius as Dy$^{3+}$ (102.7 pm), whereas Ca$^{2+}$ with an ionic radius of 112 pm is larger than Dy in size \cite{Shannon1976}. The larger size of Ca$^{2+}$ might be at the origin for its effectiveness in changing the magnetic exchange interaction and/or crystal field, and could be the reason for stronger smearing of spin-freezing anomaly in ac susceptibility than by Y$^{3+}$. A comparative study of crystal field level schemes of pure Dy$_2$Ti$_2$O$_7$ as well as Ca- and Y-substituted ones would be informative in understanding the magnetic behaviors of these compounds. In a recent study of effect of dilution on spin-ice behavior Scharffe et al.\ \cite{Scharffe2015} found evidence for suppression of Pauling’s residual entropy in Dy$_{2-x}$Y$_x$Ti$_2$O$_7$ for $x > 0.4$ with a non-degenerate ground state and accordingly a rapid crossover from spin-ice behavior in Dy$_2$Ti$_2$O$_7$ to a weakly interacting single-ion physics is inferred for intermediate dilution in Dy$_{2-x}$Y$_x$Ti$_2$O$_7$. It would be of interest to see how Ca substitution for Dy influences the spin-ice behavior and associated monopole dynamics. Further investigations in this direction would be enlightening.    

\acknowledgements
We thank B. Klemeke and K. Siemensmeyer for their experimental assistance and helpful discussions. We acknowledge Helmholtz Gemeinschaft for funding via the Helmholtz Virtual Institute (Project No. VH-VI-521).


\begin{thebibliography}{40}

\bibitem{Gardner2010}
Gardner J S, Gingras M J P and Greedan J E 2010 {\it Rev. Mod. Phys. } {\bf 82} 53

\bibitem{Castelnovo2012}
Castelnovo C , Moessner R and Sondhi S L 2012 {\it Annu. Rev. Condens. Matter Phys. } {\bf 3} 35

\bibitem{Harris1997}
Harris M J, Bramwell S T,  McMorrow D F, Zeiske T and Godfrey K W 1997  {\it Phys. Rev. Lett. } {\bf 79} 2554

\bibitem{Ramirez1999}
A. P. Ramirez, A. Hayashi, R. J. Cava, R. B. Siddharthan,  and S. Shastry, {\it Nature} {\bf  399} 333 (1999).

\bibitem{Siddharthan1999}
Siddharthan R, Shastry B S, Ramirez A P, Hayashi A, Cava R J and Rosenkranz S 1999 {\it Phys. Rev. Lett. } {\bf 83} 1854

\bibitem{Hertog2000}
Hertog B C D and Gingras M J P 2000 {\it Phys. Rev. Lett.} {\bf 84} 3430

\bibitem{Bramwell2001}
Bramwell S T and Gingras M J P 2001 {\it Science} {\bf  294} 1495

\bibitem{Castelnovo2008}
Castelnovo C, Moessner R and Sondhi S L 2008 {\it Nature} {\bf 451} 42

\bibitem{Morris2009}
Morris D J P, Tennant D A, Grigera S A, Klemke B, Castelnovo  C, Moessner R, Czternasty C, Meissner M,  Rule K C, Hoffmann J, Kiefer K, Gerischer S, Slobinsky D and Perry R S 2009 {\it Science} {\bf 326} 411

\bibitem{Bramwell2009}
Bramwell S T, Giblin S R, Calder S, Aldus R, Prabhakaran D and Fennell T 2009 {\it Nature } {\bf 461} 956

\bibitem{Fennell2009}
Fennell T, Deen P P, Wildes A R, Schmalzl K, Prabhakaran D, Boothroyd A T, Aldus R J, McMorrow D F and Bramwell S T 2009  {\it Science} {\bf  326} 415

\bibitem{Snyder2001}
Snyder J, Slusky J S, Cava R J and Schiffer P 2001 {\it Nature } {\bf 413} 48

\bibitem{Matsuhira2001}
Matsuhira K,  Hinatsu Y and Sakakibara T 2001 {\it J. Phys.: Condens. Matter } {\bf 13} L737

\bibitem{Snyder2002}
Snyder J, Slusky J S, Cava R J and Schiffer P 2002 {\it Phys. Rev. B} {\bf  66} 064432

\bibitem{Snyder2004}
Snyder J, Ueland B G, Mizel A, Slusky J S, Karunadasa H, Cava R J and Schiffer P 2004 {\it Phys. Rev. B } {\bf 70} 184431

\bibitem{Orbach1961}
Orbach R 1961 {\it Proc. R. Soc. Lond. A} {\bf  264} 458

\bibitem{Ehlers2003}
Ehlers G, Cornelius A L, Orend\'a\v{c} M, Kaj\v{n}akov\'a M, Fennell T, Bramwell S T and Gardner J S 2003 {\it J. Phys. Condens. Matter} {\bf 15} L9

\bibitem{Rodriguez1993}
J. Rodr\'{i}guez-Carvajal 1993 {\it Physica B} {\bf  192} 55; Program Fullprof, LLB-JRC, Laboratoire L\'{e}on Brillouin, CEA-Saclay, France, 1996 (www.ill.eu/sites/fullprof/).

\bibitem{Fuentes2005}
Fuentes A F, Boulahya K,  Maczka M, Hanuza J and Amador U 2005  {\it Solid State Sci.} {\bf  7} 343

\bibitem{Mydosh1993}
Mydosh J A 1993  \emph{Spin Glasses: An Experimental Introduction} (Taylor and Francis, London)

\bibitem{Bramwell2000}
Bramwell S T, Field M N, Harris M J and Parkin I P 2000  {\it J. Phys.: Condens. Matter } {\bf 12} 483

\bibitem{Jana2002}
Jana Y M, Sengupta A and Ghosh D 2002 {\it J. Magn. Magn. Mater.} {\bf  248} 7

\bibitem{Fukazawa2002}
Fukazawa H, Melko R J, Higashinaka R,  Maeno Y and Gingras M J P 2002 {\it Phys. Rev. B } {\bf 65} 054410

\bibitem{Sala2014}
Sala G, Gutmann M J, Prabhakaran D, Pomaranski D, Mitchelitis C, Kycia J B, Porter D G, Castelnovo C and Go J P 2014 {\it Nature Materials} {\bf  13} 488

\bibitem{Kittel2005}
Kittel C 2005 \emph{Introduction to Solid State Physics}, 8th ed. (Wiley, New York).

\bibitem{Gopal1966}
Gopal E S R 1966 \emph{Specific Heats at Low Temperatures} (Plenum, New York).

\bibitem{Hiroi2003}
Hiroi Z, Matsuhira K, Ogata M 2003 {\it J. Phys. Soc. Jpn. } {\bf 72} 3045

\bibitem{Klemke2011}
Klemke B, Meissner M, Strehlow P, Kiefer K, Grigera S A and Tennant D A 2011 {\it J. Low Temp. Phys.} {\bf  163} 345

\bibitem{Rosenkranz2000}
Rosenkranz S, Ramirez A P, Hayashi A, Cava R J, Siddharthan R and Shastry B S 2000 {\it J. Appl. Phys.} {\bf 87} 5914

\bibitem{Malkin2004}
Malkin B Z, Zakirov A R, Popova M N, Klimin S A, Chukalina E P,  Antic-Fidancev E, Goldner P, Aschehoug P and Dhalenne G 2004 {\it Phys. Rev. B } {\bf 70}  075112

\bibitem{Lummen2008}
Lummen T T A, Handayani I P,  Donker M C, Fausti D, Dhalenne G, Berthet P, Revcolevschi A and  van Loosdrecht P H M 2008 {\it Phys. Rev. B } {\bf 77} 214310

\bibitem{Goetsch2012}
Goetsch R J, Anand V K, Pandey A and Johnston D C 2012 {\it Phys. Rev. B} {\bf 85} 054517

\bibitem{Johnson2009}
Johnson M B, James D D, Bourque A, Dabkowska H A, Gaulin B D and White  M A 2009 {\it J. Solid State Chem.} {\bf 182} 725

\bibitem{Snyder2003}
Snyder J, Ueland B G, Slusky J S, Karunadasa H, Cava R J, Mizel A and Schiffer P 2003 {\it Phys. Rev. Lett.} {\bf 91} 107201 

\bibitem{Ehlers2004}
Ehlers G, Cornelius A L, Fennell T, Koza M, Bramwell S T and Gardner J S 2004 {\it J. Phys.: Condens. Matter} {\bf 16} S635

\bibitem{Shannon1976}
Shannon R D 1976 {\it  Acta Cryst.} {\bf A32} 751

\bibitem{Scharffe2015}
Scharffe S,  Breunig O, Cho V, Laschitzky P, Valldor M, Welter J F and Lorenz T 2015 arXiv: 1503.03856

\end{thebibliography}
\end{document}